\documentclass[a4paper,aps,twocolumn,nofootinbib]{revtex4}
\RequirePackage[colorlinks,hyperindex]{hyperref}
\RequirePackage[english]{babel}
\RequirePackage[latin1]{inputenc}
\RequirePackage[T1]{fontenc}
\RequirePackage{mathrsfs}
\RequirePackage{amsmath}
\RequirePackage{amssymb}
\RequirePackage{amsbsy}
\RequirePackage{color}
\RequirePackage{bm}
\hypersetup{colorlinks=true,breaklinks=true,urlcolor=blue,linkcolor=red}
\begin{document}
\title{\bf{Dirac field, van der Waals gas, Weyssenhoff fluid, Newton particle}}
\author{Luca Fabbri$^{\nabla}$\!\!\! $^{\hbar}$\footnote{luca.fabbri@unige.it}}
\affiliation{$^{\nabla}$DIME, Universit\`{a} di Genova, Via all'Opera Pia 15, 16145 Genova, ITALY\\
$^{\hbar}$INFN, Sezione di Genova, Via Dodecaneso 33, 16146 Genova, ITALY}
\date{\today}
\begin{abstract}
We consider the Dirac field in polar formulation, showing that when torsion is taken in effective approximation the theory has the thermodynamic properties of a van der Waals gas, that when the limit of zero chiral angle is taken the theory reduces to that of a Weyssenhoff fluid, and that under the spinless condition it gives the Newtonian particle. This nesting of approximations will allow us to interpret the various objects pertaining to the spinor, with torsion providing a form of negative pressure, and the chiral angle being related to a type of temperature.
\end{abstract}
\maketitle
\section{Introduction}
Both in geometric construction, and for its far-reaching applications, the Dirac field is among the most important fields in mathematics and physics. Still, when confronted to possible interpretations, there appears to be the spread consensus that no-one really understands what the spinor actually is. This situation is not limited to the relativistic spinor field. The Pauli field is affected by the very same condition. Neither is this situation confined to relativistic and non-relativistic spinor fields. The Schr\"{o}dinger wave function carries the same burden. So whether constituted by two chiral states or only one, whether characterized by two helicities or a single one, what seems to be at the root of the problem is the fact that all these wave functions are intrinsically built to be complex-valued fields.

On the other hand, all complex quantities may always be written in polar form, in which complex functions are re-expressed as product of modules times unitary phases, with modules and phases being real. Pauli spinors, having two helicities, need extra care in under-going the polar decomposition since, under rotations, the two components would mix. And even more care is required for Dirac spinors since, having two helicities as well as two chiralities, under Lorentz transformations all four components would mix. Still, the relativistic polar formulation is doable just as well, as was first shown in \cite{jl1, jl2}.

The advantage on the polar decomposition of relativistic spinor fields is that it converts the entire Dirac theory into a form that is genuinely hydrodynamic \cite{Fabbri:2023onb}. Clearly, this does not only mean that all variables are real. It also means that all variables are in themselves perfectly visualizable in terms of concepts of fluid mechanics. Indeed, of the four sets of variables in terms of which the spinor field can be decomposed, two are the density and velocity, exactly as those we have in hydrodynamics. Another is the spin, which has become well known in modern physics nowadays. The final one is the chiral angle, which is not yet easy to understand, although we hope that it will be better clarified in the light of the investigations that are to be done in this work. We will see that, under general conditions, the chiral angle can be interpreted as a form of generalized temperature. When chiral angle, density, spin and velocity are all accounted for, one can see that the Dirac field theory is re-formulated as a type of fluid with a temperature and a pressure verifying the relationships they would satisfy in the case of the van der Waals gas. We will also see that in the zero-temperature regime, such a gas behaves as a Weyssenhoff fluid of completely antisymmetric spin. And eventually, in spinlessness case the laws of the Newtonian dynamics are recovered.

The idea of re-formulating quantum mechanics as some type of fluid dates back to the works of Madelung, who first considered writing the wave function as a product of module and phase, respectively related to density and velocity. In turn, this would split the Schr\"{o}dinger equation into one Hamilton-Jacobi equation with a quantum potential written in terms of the density and one continuity equation for the velocity. This was the basis upon which Bohm started to build his interpretation of quantum mechanics \cite{b}. The treatment has also been revised by Takabayasi in \cite{t}. The relativistic extension has been attempted first by Bohm in \cite{b1}. And hence by Takabayasi in a series of works culminating with reference \cite{Takabayasi1957}.

All these works have in common with our present treatment the idea of trying to write relativistic quantum mechanics as a type of classical mechanics. But none could reach a fully general covariant description because they never considered the polar form first proposed in \cite{jl1, jl2}. It is our objective to show that when the polar form of \cite{jl1, jl2} is used as done in \cite{Fabbri:2023onb}, all results of Bohm and Takabayasi can find their most generally covariant expression.
\section{Dirac Field in Polar Form}
\subsection{Dirac Spinors}
We start with a brief summary of the Dirac spinors to set our convention, and to establish the relations that we are going to need later on. So to begin, let $\boldsymbol{\gamma}^{i}$ be matrices belonging to the Clifford algebra $\{\boldsymbol{\gamma}^{i},\boldsymbol{\gamma}^{j}\}\!=\!2\mathbb{I}\eta^{ij}$ with $\eta^{ij}$ the Minkowski matrix. Then $\boldsymbol{\sigma}_{ik}\!=\![\boldsymbol{\gamma}_{i},\boldsymbol{\gamma}_{k}]/4$ are defined as generators of the Lorentz group. In $2i\boldsymbol{\sigma}_{ab}\!=\!\varepsilon_{abcd}\boldsymbol{\pi}\boldsymbol{\sigma}^{cd}$ we find the implicit definition of the parity-odd matrix $\boldsymbol{\pi}$ and whose existence stipulates that the Lorentz group is reducible.\footnote{This is the fifth gamma matrix, which we will not indicate as a gamma with an index five to avoid the confusion coming from the dummy index. The Greek letter $\boldsymbol{\pi}$ corresponds to the Latin letter $p$ and it stands for \emph{parity} in the same way that the Greek letter $\boldsymbol{\sigma}$ corresponds to the Latin letter $s$ and it stands for \emph{spin}.}\! The exponentiation of the generators gives an element of the Lorentz group $\boldsymbol{\Lambda}$ and therefore $\boldsymbol{S}\!=\!\boldsymbol{\Lambda}e^{iq\alpha}$ is an element of the spinor group accounting also for gauge transformations for which $q$ is the charge. A spinor field is an object that transforms like $\psi\!\rightarrow\!\boldsymbol{S}\psi$ and $\overline{\psi}\!\rightarrow\!\overline{\psi}\boldsymbol{S}^{-1}$ where $\overline{\psi}\!=\!\psi^{\dagger}\boldsymbol{\gamma}^{0}$ is the adjoint operation. With a pair of adjoint spinors we can form the spinorial bi-linears
\begin{eqnarray}
&\Sigma^{ab}\!=\!2\overline{\psi}\boldsymbol{\sigma}^{ab}\boldsymbol{\pi}\psi\ \ \ \
\ \ \ \ \ \ \ \ M^{ab}\!=\!2i\overline{\psi}\boldsymbol{\sigma}^{ab}\psi\label{tensors}\\
&S^{a}\!=\!\overline{\psi}\boldsymbol{\gamma}^{a}\boldsymbol{\pi}\psi\ \ \ \
\ \ \ \ \ \ \ \ U^{a}\!=\!\overline{\psi}\boldsymbol{\gamma}^{a}\psi\label{vectors}\\
&\Theta\!=\!i\overline{\psi}\boldsymbol{\pi}\psi\ \ \ \
\ \ \ \ \ \ \ \ \Phi\!=\!\overline{\psi}\psi\label{scalars}
\end{eqnarray}
which are all real tensors. They verify the Hodge duality
\begin{eqnarray}
&\Sigma^{ab}\!=\!-\frac{1}{2}\varepsilon^{abij}M_{ij}\ \ \ \
\ \ \ \ M^{ab}\!=\!\frac{1}{2}\varepsilon^{abij}\Sigma_{ij}\label{dual}
\end{eqnarray}
beside the constitutive relations
\begin{eqnarray}
&M_{ik}U^{i}=\Theta S_{k}\ \ \ \ \ \ \ \ \Sigma_{ik}U^{i}\!=\!\Phi S_{k}\\
&M_{ik}S^{i}=\Theta U_{k}\ \ \ \ \ \ \ \ \Sigma_{ik}S^{i}\!=\!\Phi U_{k}
\label{products}
\end{eqnarray}
as well as
\begin{eqnarray}
&M_{ab}\Phi\!-\!\Sigma_{ab}\Theta\!=\!U^{j}S^{k}\varepsilon_{jkab}\\
&M_{ab}\Theta\!+\!\Sigma_{ab}\Phi\!=\!U_{[a}S_{b]}
\end{eqnarray}
together with
\begin{eqnarray}
&\frac{1}{2}M_{ab}M^{ab}\!=\!-\frac{1}{2}\Sigma_{ab}\Sigma^{ab}\!=\!\Phi^{2}\!-\!\Theta^{2}\\
&\frac{1}{2}M_{ab}\Sigma^{ab}\!=\!-2\Theta\Phi
\end{eqnarray}
and
\begin{eqnarray}
&U_{a}U^{a}\!=\!-S_{a}S^{a}\!=\!\Theta^{2}\!+\!\Phi^{2}\label{NORM}\\
&U_{a}S^{a}\!=\!0\label{ORTHOGONAL}
\end{eqnarray}
called Fierz re-arrangements. They show that not all the bi-linears are independent, and in fact if $\Phi^{2}\!+\!\Theta^{2}\!\neq\!0$ then both antisymmetric tensors $M_{ab}$ and $\Sigma_{ab}$ can be dropped in favour of the two vectors and the two scalars. In turn, under the same condition, the axial-vector and the vector $S_{a}$ and $U_{a}$ are space-like and time-like, showing that they can be recognized as spin and velocity, respectively \cite{Fabbri:2023onb}.

The spinorial covariant derivative is defined as
\begin{eqnarray}
&\boldsymbol{\nabla}_{\mu}\psi\!=\!\partial_{\mu}\psi
\!+\!\frac{1}{2}C_{ab\mu}\boldsymbol{\sigma}^{ab}\psi\!+\!iqA_{\mu}\psi
\label{spincovder}
\end{eqnarray}
in which $A_{\mu}$ is the gauge potential and $C^{ab}_{\phantom{ab}\mu}$ is the space-time spin connection. We are here in the torsionless case although full generality will be recovered by introducing torsion as an axial-vector field in the dynamics.

As usual, the commutator
\begin{eqnarray}
&[\boldsymbol{\nabla}_{\mu},\! \boldsymbol{\nabla}_{\nu}]\psi
\!=\!\frac{1}{2}R_{ab\mu\nu}\boldsymbol{\sigma}^{ab}\psi\!+\!iqF_{\mu\nu}\psi
\end{eqnarray}
defines the Riemann curvature and the Maxwell strength.

The dynamics is assigned by the torsion field equations
\begin{eqnarray}
&\nabla_{\rho}(\partial W)^{\rho\mu}\!+\!M^{2}W^{\mu}\!=\!XS^{\mu}
\label{torsion}
\end{eqnarray}
together with the gravitational field equations
\begin{eqnarray}
\nonumber
&R^{\rho\sigma}\!-\!\frac{1}{2}Rg^{\rho\sigma}\!-\!\Lambda g^{\rho\sigma}
\!=\!\frac{1}{2}[\frac{1}{4}F^{2}g^{\rho\sigma}
\!-\!F^{\rho\alpha}\!F^{\sigma}_{\phantom{\sigma}\alpha}+\\
\nonumber
&+\frac{1}{4}(\partial W)^{2}g^{\rho\sigma}
\!-\!(\partial W)^{\sigma\alpha}(\partial W)^{\rho}_{\phantom{\rho}\alpha}+\\
\nonumber
&+M^{2}(W^{\rho}W^{\sigma}\!-\!\frac{1}{2}W^{2}g^{\rho\sigma})+\\
\nonumber
&+\frac{i}{4}(\overline{\psi}\boldsymbol{\gamma}^{\rho}\boldsymbol{\nabla}^{\sigma}\psi
\!-\!\boldsymbol{\nabla}^{\sigma}\overline{\psi}\boldsymbol{\gamma}^{\rho}\psi
\!+\!\overline{\psi}\boldsymbol{\gamma}^{\sigma}\boldsymbol{\nabla}^{\rho}\psi
\!-\!\boldsymbol{\nabla}^{\rho}\overline{\psi}\boldsymbol{\gamma}^{\sigma}\psi)-\\
&-\frac{1}{2}X(W^{\sigma}S^{\rho}\!+\!W^{\rho}S^{\sigma})]
\label{gravitation}
\end{eqnarray}
and the electrodynamic field equations
\begin{eqnarray}
&\nabla_{\sigma}F^{\sigma\mu}\!=\!qU^{\mu}
\label{electrodynamics}
\end{eqnarray}
where $(\partial W)_{\alpha\nu}\!=\!\nabla_{\alpha}W_{\nu}\!-\!\nabla_{\nu}W_{\alpha}$ and $M$ the torsion mass, and where we define $R^{\alpha}_{\phantom{\alpha}\rho\alpha\sigma}\!=\!R_{\rho\sigma}$ and $R^{\rho\sigma}g_{\rho\sigma}\!=\!R$ as the Ricci tensor and scalar and $\Lambda$ the cosmological constant.

As for matter, the dynamics is assigned in terms of the Dirac spinor field equation given by
\begin{eqnarray}
&i\boldsymbol{\gamma}^{\mu}\boldsymbol{\nabla}_{\mu}\psi
\!-\!XW_{\sigma}\boldsymbol{\gamma}^{\sigma}\boldsymbol{\pi}\psi\!-\!m\psi\!=\!0
\label{matter}
\end{eqnarray}
where $W_{\sigma}$ is the Hodge dual of the torsion tensor and $X$ the torsion-spin coupling constant, added to recover full generality as we have already anticipated \cite{Fabbri:2023cot}.

The set of field equations (\ref{torsion}-\ref{gravitation}) with (\ref{electrodynamics}) is conceived in this way so to give rise to conservation laws that turn out to be automatically satisfied when the Dirac spinorial field equations (\ref{matter}) are valid, and so it is consistent.
\subsection{Polar Decomposition}
In the aforementioned case in which $\Phi^{2}\!+\!\Theta^{2}\!\neq\!0$ we can perform what is called polar decomposition of the spinor field. Specifically, it is possible to demonstrate \cite{jl1,jl2} that under the above condition any spinor field can always be written, in chiral representation, in the form
\begin{eqnarray}
&\psi\!=\!\phi\ e^{-\frac{i}{2}\beta\boldsymbol{\pi}}
\ \boldsymbol{L}^{-1}\left(\begin{tabular}{c}
$1$\\
$0$\\
$1$\\
$0$
\end{tabular}\right)
\label{spinor}
\end{eqnarray}
for a pair of functions $\phi$ and $\beta$ and for some $\boldsymbol{L}$ with the structure of a spinor transformation. As anticipated, the two antisymmetric tensors are expressed by means of the two vectors and the two scalars, and these are given by
\begin{eqnarray}
&S^{a}\!=\!2\phi^{2}s^{a}\ \ \ \
\ \ \ \ \ \ \ \ U^{a}\!=\!2\phi^{2}u^{a}
\end{eqnarray}
and
\begin{eqnarray}
&\Theta\!=\!2\phi^{2}\sin{\beta}\ \ \ \
\ \ \ \ \ \ \ \ \Phi\!=\!2\phi^{2}\cos{\beta}
\end{eqnarray}
when the polar form is implemented. The last two show that $\phi$ and $\beta$ are a scalar and a pseudo-scalar, known as module and chiral angle. Then (\ref{NORM}-\ref{ORTHOGONAL}) reduce to
\begin{eqnarray}
&u_{a}u^{a}\!=\!-s_{a}s^{a}\!=\!1\label{norm}\\
&u_{a}s^{a}\!=\!0\label{orthogonal}
\end{eqnarray}
showing that the velocity has only $3$ independent components, the $3$ spatial rapidities, whereas the spin has only $2$ independent components, the $2$ angles that, in the rest-frame, its spatial part forms with the third axis. As for $\boldsymbol{L}$ we can read its meaning as that of the specific transformation that takes a given spinor to its rest-frame with spin aligned along the third axis. For the spinorial fields in polar form, the $8$ real components are re-configured in such a way that the $2$ scalars $\phi$ and $\beta$ are isolated from the $6$ parameters of $\boldsymbol{L}$ that can always be transferred into the frame and which are thus the Goldstone fields.

Because in general one can prove that
\begin{eqnarray}
&\boldsymbol{L}^{-1}\partial_{\mu}\boldsymbol{L}\!=\!iq\partial_{\mu}\zeta\mathbb{I}
\!+\!\frac{1}{2}\partial_{\mu}\zeta_{ij}\boldsymbol{\sigma}^{ij}\label{spintrans}
\end{eqnarray}
for some $\zeta$ and $\zeta_{ij}$ then we can define
\begin{eqnarray}
&R_{ij\mu}\!:=\!\partial_{\mu}\zeta_{ij}\!-\!C_{ij\mu}\label{R}\\
&P_{\mu}\!:=\!q(\partial_{\mu}\zeta\!-\!A_{\mu})\label{P}
\end{eqnarray}
which are real tensors. By reading these expressions one can see that after the Goldstone fields are transferred into the frame, they combine with spin connection and gauge potential to become the longitudinal components of the $P_{\mu}$ and $R_{ij\mu}$ tensors, hence called gauge and space-time tensorial connections. From (\ref{spinor}) with (\ref{P}-\ref{R}) we get
\begin{eqnarray}
&\!\!\!\!\!\!\!\!\boldsymbol{\nabla}_{\mu}\psi\!=\!(\nabla_{\mu}\ln{\phi}\mathbb{I}
\!-\!\frac{i}{2}\nabla_{\mu}\beta\boldsymbol{\pi}
\!-\!\frac{1}{2}R_{\alpha\nu\mu}\boldsymbol{\sigma}^{\alpha\nu}
\!-\!iP_{\mu}\mathbb{I})\psi
\label{decspinder}
\end{eqnarray}
as the polar form of the covariant derivative. Notice that
\begin{eqnarray}
&\nabla_{\mu}s_{\nu}\!=\!s^{\alpha}R_{\alpha\nu\mu}\ \ \ \
\ \ \ \ \nabla_{\mu}u_{\nu}\!=\!u^{\alpha}R_{\alpha\nu\mu}\label{ds-du}
\end{eqnarray}
as general identities. The covariant derivative of the velocity is the object with which one builds the strain-rate tensor in continuum mechanics. Expressions (\ref{ds-du}) are the extension to both velocity and spin of relationships that make $R_{ab\mu}$ interpretable as the strain-rate tensor.

The tensorial connections are such that
\begin{eqnarray}
&\!\!\!\!\!\!\!\!-R^{i}_{\phantom{i}j\mu\nu}\!=\!\nabla_{\mu}R^{i}_{\phantom{i}j\nu}
\!-\!\nabla_{\nu}R^{i}_{\phantom{i}j\mu}\!+\!R^{i}_{\phantom{i}k\mu}R^{k}_{\phantom{k}j\nu}
\!-\!R^{i}_{\phantom{i}k\nu}R^{k}_{\phantom{k}j\mu}\\
&\!\!\!\!-qF_{\mu\nu}\!=\!\nabla_{\mu}P_{\nu}\!-\!\nabla_{\nu}P_{\mu}
\end{eqnarray}
therefore being the covariant potentials of these tensors.

In the gravitational field equations, the right-hand side aside for the factor $1/2$ is the energy density tensor, and it is expressed in polar variables according to
\begin{eqnarray}
\nonumber
&T^{\rho\sigma}\!=\!\frac{1}{4}F^{2}g^{\rho\sigma}
\!-\!F^{\rho\alpha}\!F^{\sigma}_{\phantom{\sigma}\alpha}+\\
\nonumber
&+\frac{1}{4}(\partial W)^{2}g^{\rho\sigma}
\!-\!(\partial W)^{\sigma\alpha}(\partial W)^{\rho}_{\phantom{\rho}\alpha}+\\
\nonumber
&+M^{2}(W^{\rho}W^{\sigma}\!-\!\frac{1}{2}W^{2}g^{\rho\sigma})+\\
\nonumber
&+\phi^{2}[P^{\rho}u^{\sigma}\!+\!P^{\sigma}u^{\rho}+\\
\nonumber
&+(\nabla^{\rho}\beta/2\!-\!XW^{\rho})s^{\sigma}
\!+\!(\nabla^{\sigma}\beta/2\!-\!XW^{\sigma})s^{\rho}-\\
&-\frac{1}{4}R_{\alpha\nu}^{\phantom{\alpha\nu}\sigma}s_{\kappa}
\varepsilon^{\rho\alpha\nu\kappa}
\!-\!\frac{1}{4}R_{\alpha\nu}^{\phantom{\alpha\nu}\rho}s_{\kappa}
\varepsilon^{\sigma\alpha\nu\kappa}]
\label{energy}
\end{eqnarray}
in terms of the space-time tensorial connection.

The Dirac spinor field equations in polar form are
\begin{eqnarray}
&\nabla_{\mu}\beta\!-\!2XW_{\mu}\!+\!B_{\mu}
\!-\!2P^{\iota}u_{[\iota}s_{\mu]}\!+\!2ms_{\mu}\cos{\beta}\!=\!0
\label{dep1}\\
&\nabla_{\mu}\ln{\phi^{2}}\!+\!R_{\mu}
\!-\!2P^{\rho}u^{\nu}s^{\alpha}\varepsilon_{\mu\rho\nu\alpha}\!+\!2ms_{\mu}\sin{\beta}\!=\!0
\label{dep2}
\end{eqnarray}
in which $R_{\mu\nu}^{\phantom{\mu\nu}\nu}\!=\!R_{\mu}$ and $\frac{1}{2}\varepsilon_{\mu\alpha\nu\iota}R^{\alpha\nu\iota}\!=\!B_{\mu}$ were defined.

Upon the introduction of the potentials
\begin{eqnarray}
&2Y_{\mu}\!=\!\nabla_{\mu}\beta\!-\!2XW_{\mu}\!+\!B_{\mu}\label{Y}\\
&2Z_{\mu}\!=\!\nabla_{\mu}\ln{\phi^{2}}\!+\!R_{\mu}\label{Z}
\end{eqnarray}
it becomes easier to work the polar spinor field equations (\ref{dep1}-\ref{dep2}) in order to isolate the gauge tensorial connection
\begin{eqnarray}
&P^{\eta}\!=\!m\cos{\beta}u^{\eta}\!+\!Y_{\mu}u^{[\mu}s^{\eta]}
\!+\!Z_{\mu}u_{\pi}s_{\tau}\varepsilon^{\mu\pi\tau\eta}\label{momentum}
\end{eqnarray}
which is recognized to be the momentum of the field and with which the energy (\ref{energy}) acquires the form
\begin{eqnarray}
\nonumber
&T^{\rho\sigma}\!=\!\frac{1}{4}F^{2}g^{\rho\sigma}
\!-\!F^{\rho\alpha}\!F^{\sigma}_{\phantom{\sigma}\alpha}+\\
\nonumber
&+\frac{1}{4}(\partial W)^{2}g^{\rho\sigma}
\!-\!(\partial W)^{\sigma\alpha}(\partial W)^{\rho}_{\phantom{\rho}\alpha}+\\
\nonumber
&+M^{2}(W^{\rho}W^{\sigma}\!-\!\frac{1}{2}W^{2}g^{\rho\sigma})+\\
\nonumber
&+\phi^{2}[2m\cos{\beta}u^{\rho}u^{\sigma}-\\
\nonumber
&-2Y_{\mu}s^{\mu}u^{\rho}u^{\sigma}
\!+\!Y_{\mu}u^{\mu}(s^{\rho}u^{\sigma}\!+\!s^{\sigma}u^{\rho})
\!+\!Y^{\rho}s^{\sigma}\!+\!Y^{\sigma}s^{\rho}+\\
\nonumber
&+Z_{\mu}u_{\pi}s_{\tau}(\varepsilon^{\mu\pi\tau\sigma}u^{\rho}
\!+\!\varepsilon^{\mu\pi\tau\rho}u^{\sigma})-\\
\nonumber
&-\frac{1}{4}(R_{\alpha\nu\pi}\varepsilon^{\rho\alpha\nu\pi}g^{\sigma\kappa}
\!+\!R_{\alpha\nu\pi}\varepsilon^{\sigma\alpha\nu\pi}g^{\rho\kappa}+\\
&+R_{\alpha\nu}^{\phantom{\alpha\nu}\sigma}\varepsilon^{\rho\alpha\nu\kappa}
\!+\!R_{\alpha\nu}^{\phantom{\alpha\nu}\rho}\varepsilon^{\sigma\alpha\nu\kappa})s_{\kappa}]
\label{energymomentum}
\end{eqnarray}
in terms of the $R_{ab\mu}$ tensor and the $Y_{\mu}$ and $Z_{\mu}$ potentials.
\section{Torsion Effective Approximation and van der Waals gas}
\subsection{General Thermodynamic Variables}
When combining the two principles of thermodynamics into the single relation $dU\!=\!TdS\!-\!pdV$ and considering that $dS$ is an exact differential form, one can extract
\begin{eqnarray}
\left(\frac{\partial U}{\partial V}\right)_{T}\!\!\!=\!
\left(\frac{\partial p}{\partial T}\right)_{V}\!\!\!\!T\!-\!p
\label{U}
\end{eqnarray}
in case $V$ and $T$ are the independent variables.

With this equation, which is usually known as internal energy equation, one can deduce the internal energy once the equation of state is assigned. For example, the simplest non-perfect gas, the van der Waals gas, has equation
\begin{eqnarray}
\left(p\!+\!\frac{a}{V^{2}}\right)(V\!-\!b)\!=\!RT
\end{eqnarray}
in which $a$ is a constant related to the effective pressure due to forces between the molecules, positive in the case of attraction, and $b$ is the volume that is occupied by the molecules. By means of \eqref{U} one can deduce that
\begin{eqnarray}
U\!=\!\omega\!+\!C_{V}T\!-\!\frac{a}{V}
\end{eqnarray}
where $\omega$ is a generic constant.

With the equation of the internal energy we are giving an axiomatic definition of thermodynamical variables in the sense that we are assigning a meaning to the different terms entering \eqref{U} according to the role they play in such an equation. For example, if we knew that $U$ had a given dependence on $V$ then the right-hand side of \eqref{U} would be known, and any pair of variables satisfying the right-hand side of \eqref{U} in exactly the way $p$ and $T$ are would respectively be interpreted as pressure and temperature.

With this in mind, we are now going to investigate the thermodynamic structure of the Dirac spinor field theory.
\subsection{Massive Propagating Torsion}
We will consider the Dirac theory with torsion taken to be massive enough to allow the effective approximation.

In effective approximation, the torsion field loses all its propagating properties, with field equations reducing to
\begin{eqnarray}
&M^{2}W^{\mu}\!=\!XS^{\mu}
\label{eff}
\end{eqnarray}
so that now torsion can be replaced in terms of the spin.

When this is done in the expression of the energy density tensor (\ref{energymomentum}) remarkable simplifications occur. Taking in particular the purely spinorial contribution, it reads
\begin{eqnarray}
\nonumber
&E^{\rho\sigma}\!=\!\phi^{2}[2(m\cos{\beta}\!-\!\phi^{2}X^{2}/M^{2}-\\
\nonumber
&-s^{\mu}\nabla_{\mu}\beta/2
\!-\!\frac{1}{4}R^{\pi\tau\eta}s^{\kappa}\varepsilon_{\kappa\pi\tau\eta})u^{\rho}u^{\sigma}+\\
\nonumber
&+2\phi^{2}X^{2}/M^{2}(g^{\rho\sigma}\!-\!u^{\rho}u^{\sigma})+\\
\nonumber
&+(s^{\rho}u^{\sigma}\!+\!s^{\sigma}u^{\rho})u^{\mu}\nabla_{\mu}\beta/2+\\
\nonumber
&+s^{\rho}\nabla^{\sigma}\beta/2\!+\!s^{\sigma}\nabla^{\rho}\beta/2+\\
\nonumber
&+Z_{\mu}u_{\pi}s_{\tau}(\varepsilon^{\mu\pi\tau\sigma}u^{\rho}
\!+\!\varepsilon^{\mu\pi\tau\rho}u^{\sigma})-\\
\nonumber
&-\frac{1}{4}R_{\pi\tau\eta}s_{\kappa}(\varepsilon^{\rho\pi\tau\kappa}g^{\sigma\eta}
\!+\!\varepsilon^{\sigma\pi\tau\kappa}g^{\rho\eta}+\\
&+\varepsilon^{\pi\tau\eta\mu}u_{\mu}u^{\sigma}g^{\rho\kappa}
\!+\!\varepsilon^{\pi\tau\eta\mu}u_{\mu}u^{\rho}g^{\sigma\kappa})]
\label{spinorenergymomentumeff}
\end{eqnarray}
which can be worked out in detail in the following way.

Defining the quantities given by
\begin{eqnarray}
&\mu\!=\!E_{\rho\sigma}u^{\rho}u^{\sigma}\\
&p\!=\!-\frac{1}{3}E_{\rho\sigma}(g^{\rho\sigma}\!-\!u^{\rho}u^{\sigma})
\end{eqnarray}
it is easy to prove that
\begin{eqnarray}
&E^{\rho\sigma}\!=\!\mu u^{\rho}u^{\sigma}\!-\!p(g^{\rho\sigma}\!-\!u^{\rho}u^{\sigma})
\!+\!\Pi^{\rho\sigma}
\end{eqnarray}
for some $\Pi^{\rho\sigma}$ in general. Rewriting the energy density in this structure helps identifying the quantities $\mu$ and $p$ as the internal energy density and the pressure of the field.

As a consequence, in our case we have that
\begin{eqnarray}
\nonumber
&\mu\!=\!2\phi^{2}(m\cos{\beta}\!-\!\phi^{2}X^{2}/M^{2})-\\
&-[2\phi^{2}(s^{\mu}\nabla_{\mu}\beta/2
\!-\!\frac{1}{2}\varepsilon^{\kappa\alpha\mu\nu}s_{\kappa}u_{\alpha}\nabla_{\mu}u_{\nu})]\\
\nonumber
&p\!=\!-2\phi^{4}X^{2}/M^{2}-\\
&-\frac{1}{3}[2\phi^{2}(s^{\mu}\nabla_{\mu}\beta/2
\!-\!\frac{1}{2}\varepsilon^{\kappa\alpha\mu\nu}s_{\kappa}u_{\alpha}\nabla_{\mu}u_{\nu})]
\end{eqnarray}
are the internal energy density and pressure of spinors.

Introducing $2\phi^{2}\!=\!1/V$ and $U\!=\!\mu V$ they become
\begin{gather}
U\!=\!m\cos{\beta}\!+\!3RT\!-\!\frac{X^{2}}{2M^{2}}\frac{1}{V}\label{vdWU}\\
\left(p\!+\!\frac{X^{2}}{2M^{2}}\frac{1}{V^{2}}\right)V\!=\!RT\label{vdWeq}
\end{gather}
in which
\begin{eqnarray}
&3RT\!=\!-s^{\mu}\nabla_{\mu}\beta/2
\!+\!\frac{1}{2}\varepsilon^{\kappa\alpha\mu\nu}s_{\kappa}u_{\alpha}\nabla_{\mu}u_{\nu}
\label{T}
\end{eqnarray}
has also been defined. Notice that (\ref{vdWeq}) is exactly the van der Waals equation of state in the case in which $b\!=\!0$ and $2a\!=\!X^{2}/M^{2}$ showing that the torsional effective force is indeed attractive. Also notice that (\ref{vdWU}) can be recognized as the van der Waals gas internal energy if $C_{V}\!=\!3R$ and $m\!=\!\omega$ as it happens for small values of the chiral angle.

The validity of (\ref{T}) can be interpreted as the definition of temperature for the Dirac field, and it can be read as the fact that the internal dynamics of the Dirac field gets contributions from its chiral angle and its vorticity. It is not surprising that the chiral angle, the phase difference between the chiral parts, be tied to the internal dynamics and so thermodynamically associated to temperature.

We recall to the reader that the association of the chiral angle to temperature, while justified by an interpretation employing the concept of internal dynamics, is only the axiomatic type of connection in the sense explained here above. The definition of temperature given by means of the internal energy and its equation \eqref{U} is formal and not functional. We have defined $T$ according to (\ref{T}) with the aim of rendering (\ref{U}) satisfied but $T$ does not represent a chaotic motion of particles as it does in the kinetic theory.

The definition of temperature as given by \eqref{T} seems to us the only way to define something conceptually close to the idea of temperature even for systems that are not constituted by randomly distributed particles.
\section{Zero Chiral Angle and Weyssenhoff Fluid}
\subsection{Non-Relativistic Regime}
In \cite{Fabbri:2023onb} and references therein, we have discussed the idea of non-relativistic limit as the regime for which
\begin{gather}
\vec{u}\!\rightarrow\!0\ \ \ \ \ \ \ \ \ \ \ \ \beta\!\rightarrow\!0\label{nonrel}
\end{gather}
characterizing the difference between the two conditions in the fact that, while the first represents the lost motion, the second represents the loss of the dynamical properties that would remain even in rest-frame, thus the intrinsic, internal dynamics. This fits well in the discussion above, where it is even more reasonably justified the fact that, in non-relativistic regime, the temperature (\ref{T}) would lose all contributions coming from the material distribution.

Therefore, while the pair of conditions (\ref{nonrel}) are the non-relativistic limit, the single condition $\beta\!=\!0$ is considered as internal triviality. Or in other words, when the chiral angle vanishes we lose the internal dynamics. This is also reasonable if we consider that $\beta\!=\!0$ means no difference between the two chiral parts. Or that the zitterbewegung effect vanishes, as it was discusses in references \cite{SR, RS}.

The condition of internal triviality has also the advantage of being covariant, so it does make sense to see what is going to happen when it is consistently assumed.
\subsection{Hydrodynamics with Spin}
Assuming $\beta\!=\!0$ from the start implies that the bi-linear pseudo-scalar $\Theta\!=\!0$ identically. Hence
\begin{eqnarray}
&M_{ik}u^{i}=0\ \ \ \ \ \ \ \ \Sigma_{ik}u^{i}\!=\!2\phi^{2}s_{k}\\
&M_{ik}s^{i}=0\ \ \ \ \ \ \ \ \Sigma_{ik}s^{i}\!=\!2\phi^{2}u_{k}
\label{productsreduced}
\end{eqnarray}
alongside to
\begin{eqnarray}
&M_{ab}\!=\!2\phi^{2}u^{j}s^{k}\varepsilon_{jkab}\\
&\Sigma_{ab}\!=\!2\phi^{2}u_{[a}s_{b]}
\end{eqnarray}
and
\begin{eqnarray}
&M_{ab}M^{ab}\!=\!-\Sigma_{ab}\Sigma^{ab}\!=\!8\phi^{4}\\
&M_{ab}\Sigma^{ab}\!=\!0
\end{eqnarray}
as Fierz identities. By employing \eqref{dual} into (\ref{productsreduced}) one has
\begin{eqnarray}
&M_{ik}u^{i}=0\ \ \ \ \ \ \ \ \frac{1}{2}\varepsilon_{kiab}M^{ab}u^{i}\!=\!2\phi^{2}s_{k}\\
&M_{ik}s^{i}=0\ \ \ \ \ \ \ \ \frac{1}{2}\varepsilon_{kiab}M^{ab}s^{i}\!=\!2\phi^{2}u_{k}
\end{eqnarray}
so that focusing in particular on the first, we can re-write the two expressions according to
\begin{eqnarray}
&M^{ki}u_{i}=0\label{conv}\\
&M^{[ab}u^{c]}\!=\!\varepsilon^{abck}S_{k}\label{compl}
\end{eqnarray}
telling that the momentum is orthogonal to the velocity and that the completely antisymmetric part of the object $M^{ij}u^{k}$ is the Hodge dual of the spin axial-vector.

As a consequence of this fact, the momentum $M^{ki}$ has all the properties needed to be identified with the fundamental spin tensor of a Weyssenhoff fluid \cite{HKH,OK}.

In fact, $M^{ki}$ is antisymmetric in its indices, condition (\ref{conv}) is just the constitutive condition of the Weyssenhoff fluid while condition (\ref{compl}) is the link between spin tensor and spin axial-vector of the Weyssenhoff fluid. The only difference with a general Weyssenhoff fluid is that in our case the spin is completely antisymmetric. However, this is expected as the Dirac spinor has a completely antisymmetric spin and it is only this part that can be excited.
\section{Spinlessness and Newton Mechanics}
\subsection{Classical Limit}
At last, we discuss the case of spinlessness. Such a case is obtained in the approximation
\begin{gather}
s^{a}\!\rightarrow\!0\label{s}
\end{gather}
and it means that we are losing quantum effects. Indeed, if we were not to choose natural units, the spin would be seen to be proportional to $\hbar$ and the limit $\hbar\!\rightarrow\!0$ is what would give rise to the classical approximation condition.

Notice also that the validity of the Dirac equation gives
\begin{gather}
\nabla_{i}S^{i}\!=\!2m\Theta
\end{gather}
showing that $\beta\!\rightarrow\!0$ is implied by $S^{i}\!\rightarrow\!0$ and stating that there can be no chirality if there is no helicity.

The present limit is therefore compatible with the limit that we discussed in the previous section.
\subsection{Point Particle}
Let us then re-consider the momentum (\ref{momentum}) as well as the energy density tensor (\ref{energymomentum}) in effective approximation and in this limit. We have
\begin{eqnarray}
&P^{\eta}\!=\!(m\!-\!2\phi^{2}X^{2}/M^{2})u^{\eta}
\end{eqnarray}
and
\begin{eqnarray}
\nonumber
&T^{\rho\sigma}\!=\!\frac{1}{4}F^{2}g^{\rho\sigma}
\!-\!F^{\rho\alpha}\!F^{\sigma}_{\phantom{\sigma}\alpha}+\\
&+2\phi^{2}(m\!-\!2\phi^{2}X^{2}/M^{2})u^{\rho}u^{\sigma}
\!+\!2\phi^{4}X^{2}/M^{2}g^{\rho\sigma}
\end{eqnarray}
which next we discuss in view of their conservation laws.

To this purpose set $2\phi^{2}\!=\!\rho$ being $\rho$ the density distribution of the material field. The above become
\begin{eqnarray}
&P^{\eta}\!=\!(m\!-\!\rho X^{2}/M^{2})u^{\eta}\label{mom}
\end{eqnarray}
and
\begin{eqnarray}
\nonumber
&T^{\rho\sigma}\!=\!\frac{1}{4}F^{2}g^{\rho\sigma}
\!-\!F^{\rho\alpha}\!F^{\sigma}_{\phantom{\sigma}\alpha}+\\
&+\rho(m\!-\!\rho X^{2}/M^{2})u^{\rho}u^{\sigma}
\!+\!\frac{1}{2}\rho^{2}X^{2}/M^{2}g^{\rho\sigma}
\label{ene}
\end{eqnarray}
and for them we know that
\begin{eqnarray}
&\nabla_{\alpha}T^{\alpha\nu}\!=\!0\label{conene}
\end{eqnarray}
and
\begin{eqnarray}
&\nabla_{\alpha}(\rho u^{\alpha})\!=\!0\label{conmom}
\end{eqnarray}
must be valid as a consequence of the Dirac spinorial field equations. Taking (\ref{mom}) into (\ref{ene}) and the result into (\ref{conene}) and then employing (\ref{conmom}) we arrive at
\begin{eqnarray}
\nonumber
&\frac{1}{2}F_{\alpha\pi}\nabla^{\sigma}F^{\alpha\pi}
\!+\!F_{\alpha\pi}\nabla^{\pi}F^{\sigma\alpha}
\!-\!\nabla^{\eta}F_{\eta\alpha}F^{\sigma\alpha}+\\
&+\rho u^{\nu}\nabla_{\nu}P^{\sigma}\!-\!\nabla^{\sigma}p\!=\!0
\end{eqnarray}
where the pressure $p\!=\!-\frac{1}{2}\rho^{2}X^{2}/M^{2}$ was used.

By employing now the Maxwell equations (\ref{electrodynamics}) we get
\begin{eqnarray}
&\rho u^{\nu}\nabla_{\nu}P^{\sigma}\!=\!\nabla^{\sigma}p\!+\!q\rho F^{\sigma\alpha}u_{\alpha}
\end{eqnarray}
which is the Newton equation of hydrodynamic motion.

In total absence of torsion, no pressure remains so that it becomes possible to simplify the density on both sides and we reduce to the final
\begin{eqnarray}
&u^{\nu}\nabla_{\nu}P^{\sigma}\!=\!qF^{\sigma\alpha}u_{\alpha}
\end{eqnarray}
as the Newton equation for the motion of material points.

It is important to remark that the Newton law has been obtained without any assumption on localization for the matter distribution. With this we do not mean to imply that matter distributions cannot be localized, but rather that there is no need for this assumption at this stage.
\section{Conclusion}
In this work, we have considered the Dirac spinor field theory re-formulated in terms of the polar variables given by the $\phi$ and $\beta$ scalars with the $u_{a}$ and $s_{a}$ vectors. After conversion, the full relativistic quantum mechanics turns into a type of hydrodynamics in which $2\phi^{2}$ is the density distribution and $\beta$ the chiral angle while $u_{a}$ is the velocity and $s_{a}$ is the spin. This hydrodynamics is, therefore, an extension of the usual one since not only the density and velocity, but also the chiral angle and spin are present.

However, the general construction can be restricted to the standard hydrodynamics by removing these two extra variables. The general theory, with torsion in its effective approximation, has the same thermodynamic features of a van der Waals gas, with van der Waals pressure due to torsion, always negative since torsion is always attractive, and with temperature and internal energy being tied to the chiral angle. In the limit $\beta\!\rightarrow\!0$ (corresponding to the requirement of losing the phase difference between chiral parts) the general theory reduces to that of a Weyssenhoff fluid with completely antisymmetric spin. And for $s_{a}\!\rightarrow\!0$ (corresponding to the condition of non-quantum limit) it reduces to a Newton fluid in presence of pressure due to torsion. By vanishing torsion the usual Newton equation for the motion of material points is eventually recovered.

Aside from allowing us to see that torsion is a form of pressure and that the chiral angle can be interpreted like a type of temperature, the polar re-formulation of spinors allows the relativistic quantum mechanics to convert into a specific hydrodynamics, whose variables may perfectly be visualized, and because of this, better understood.

The challenges of relativistic quantum mechanics have no resolution in a re-formulation of the theory alone, and many questions remain still open. Nonetheless, questions that can be answered more easily when made clearer will receive a boost by a Dirac spinor field theory formulated in terms of variables that are visualizable.

In its polar form, the Dirac theory is precisely this.

\

\textbf{Funding and acknowledgements}. This work was funded by Next Generation EU project ``Geometrical and Topological effects on Quantum Matter (GeTOnQuaM)''.

\end{document}